\documentclass[useAMS,usenatbib]{tellus}
\usepackage[a4]{crop}
\usepackage{amsmath}
\usepackage{amsfonts} 
\usepackage{amssymb}
\usepackage[mathscr]{eucal}
\usepackage[dvips]{graphicx}
\usepackage{soul}
\usepackage{fancyhdr}
\usepackage{multicol}

\graphicspath{{../fig/}{./fig/}{../txt/fig/}}
\usepackage{tikz}
\usetikzlibrary{fit,shapes,arrows}
\usepackage{authblk}%affiliations
\usepackage{etoolbox}
\makeatletter
\patchcmd{\@verbatim}
  {\verbatim@font}
  {\verbatim@font\scriptsize}
  {}{}
\makeatother

\usepackage{widetext}
\makeatletter
  \newcommand\figcaption{\def\@captype{figure}\caption}
  \newcommand\tabcaption{\def\@captype{table}\caption}
\makeatother
\def\mi{\begin{equation}}
\def\mf{\end{equation}}
\def\mia{\begin{eqnarray}}
\def\mfa{\end{eqnarray}}
\def\mmi{\begin{multline}}%long equations
\usepackage{color,soul}
\usepackage[normalem]{ulem}

\def\reff#1{(\ref{#1})}

\def\mean{\overline}

\def\mrm{\mathrm}

\newcommand{\ud}{\mathrm{d}}

\renewcommand{\v}[1]{\ensuremath{\mathbf{#1}}} % for vectors
\newcommand{\gv}[1]{\ensuremath{\mbox{\boldmath$ #1 $}}}  % for vectors of Greek letters
\renewcommand{\d}[2]{\frac{\mrm{d} #1}{\mrm{d} #2}} % for derivatives
\let\baraccent=\= % rename builtin command \= to \baraccent
\renewcommand{\=}[1]{\stackrel{#1}{=}} % for putting numbers above =

\newcommand{\expec}[1]{\mathcal{E}\left( #1 \right)}
\newcommand{\esp}[1]{\mathcal{E}\left(\left. #1 \right|\v y_{1:K}\right)} % esperanza
\newcommand{\mdl}[1]{\mathcal{M}\left( #1 \right)} % modelo
\renewcommand{\H}[1]{\mathcal{H}\left( #1 \right)} %observational oper
\newcommand{\transp}{\mathrm{T}}

\renewcommand{\baselinestretch}{1.2}
\begin{document}
%\preprint{}

%\title{Stochastic parameter estimation in geophysical systems using expectation-maximization and Newton-Raphson maximum likelihood methods}
\title[Stochastic parameterization identification using maximum likelihood methods]{Stochastic parameterization identification using ensemble Kalman filtering combined with expectation-maximization and Newton-Raphson maximum likelihood methods}



%\author{Manuel Pulido}
%\affiliation{Department of Physics, FaCENA, Universidad Nacional del Nordeste, Corrientes, and IFAECI, CNRS-CONICET, Buenos Aires, Argentina.}
%\email{pulido@unne.edu.ar}
%\author{Pierre Tandeo}
%\affiliation{Institut Mines-Telecom Atlantique, UMR CNRS 6285 Lab-STICC, Brest, France}
%\author{Marc Bocquet}
%\affiliation{CEREA, joint laboratory \'Ecole des Ponts ParisTech and EDF R\&D, Universit\'e Paris-Est, Champs-sur-Marne, France}
%
%\author{Alberto Carrassi}
%\affiliation{Nansen Environmental and Remote Sensing Center, Bergen, Norway}
%\author{Magdalena Lucini}
%\affiliation{Department of Mathematics, FaCENA, Universidad Nacional del Nordeste and CONICET, Corrientes, Argentina.}

\author[1]{Manuel Pulido}
\affil[1]{\small Department of Physics, FaCENA, Universidad Nacional del Nordeste, Corrientes, and IFAECI, CNRS-CONICET, Buenos Aires, Argentina\thanks{Corresponding author.\hfil\break e-mail: pulido@unne.edu.ar}}
\author[2]{Pierre Tandeo}
\affil[2]{\small Institut Mines-Telecom Atlantique, UMR CNRS 6285 Lab-STICC, Brest, France}
\author[3]{Marc Bocquet}
\affil[3]{\small CEREA, joint laboratory \'Ecole des Ponts ParisTech and EDF R\&D, Universit\'e Paris-Est, Champs-sur-Marne, France}
\author[4]{Alberto Carrassi}
\affil[4]{\small Nansen Environmental and Remote Sensing Center, Bergen, Norway}
\author[5]{Magdalena Lucini}
\affil[5]{\small Department of Mathematics, FaCENA, Universidad Nacional del Nordeste and CONICET, Corrientes, Argentina}

\history{Manuscript received xx xxxx xx; in final form xx xxxx xx}

\maketitle

%\date{\today}

\begin{abstract}
For modelling geophysical systems, large-scale processes are described through a set of coarse-grained dynamical equations while small-scale processes are represented via parameterizations. This work proposes a method for identifying the best possible stochastic parameterization from noisy data.
State-the-art sequential estimation methods such as Kalman and particle filters do not achieve this goal succesfully because both suffer from the collapse of the parameter posterior distribution. To overcome this intrinsic limitation,  we propose two statistical learning methods. They are based on the combination of two methodologies: the maximization of the likelihood via Expectation-Maximization (EM) and Newton-Raphson (NR)  algorithms which are mainly applied in the statistic and machine learning communities, and the ensemble Kalman filter (EnKF). The methods are derived using a Bayesian  approach for a hidden Markov model. They are applied to infer deterministic and stochastic physical parameters from noisy observations in coarse-grained dynamical models.  Numerical experiments are conducted using the Lorenz-96 dynamical system with one and two scales as a proof-of-concept.  The imperfect coarse-grained model is modelled through a one-scale Lorenz-96 system in which  a stochastic parameterization is incorpored to represent the small-scale dynamics.  The algorithms are able to identify an optimal stochastic parameterization with a good  accuracy under moderate observational noise. The proposed EnKF-EM and EnKF-NR are promising statistical learning methods for developing stochastic parameterizations in high-dimensional geophysical models. 

\end{abstract}

%Popular Summary.

%The  statistical combination of observations with model data through sequential data assimilation methods allows the estimation of the model state and also model parameters. This work proposes two statistical learning methods for system identification through  stochastic parameterizations from noisy data.They are based on the combination of two independent methodologies, each one which its own well-established community: the Expectation-Maximization (EM) and Newton-Raphson (NR) maximum likelihood algorithms which are mainly applied in the statistic and machine learning communities, and the ensemble Kalman filter (EnKF) which has been widely used for geophysical applications although its use has been spreading to other applications. The methods are succesfully used to identify an optimal stochastic parameterization  using noisy observations from a multi-scale dynamical system. Multi-scale nonlinear dynamical systems are ubiquitous in biology, neurosciences and geosciences, among others. The proposed EnKF-EM and EnKF-NR are statistical learning methods that can be used in all these disciplines to improve the representation of the complex interactions between different scales. 

\section{Introduction}

The statistical combination of observations of a dynamical model with a priori information of physical laws  allows the estimation of the full state of the model even when it is only partially observed. This is the main aim of data assimilation \citep{kalnay02}. One common challenge of evolving multi-scale systems in applications ranging from meteorology, oceanography, hydrology and space physics to biochemistry and biological systems is the presence of parameters that do not rely on known physical constants so that their values are unknown and unconstrained. Data assimilation techniques can also be formulated to estimate these model parameters from observations \citep{jazwinski70,wikle07}.

There are several multi-scale physical systems which are modelled through coarse-grained equations.  The most paradigmatic cases being climate models \citep{stensrud09}, large-eddy simulations of turbulent flows \citep{mason92}, and electron fluxes in the radiation belts \citep{kondrashov11}. These imperfect models need to include  subgrid-scale effects through physical parameterizations \citep{nicolis04}. In the last years, stochastic physical parameterizations have been incorporated in weather forecast and climate models \citep{palmer01,shutts15,christensen15}. They are called stochastic parameterizations because they represent stochastically a process that is not explicitly resolved in the model, even when the unresolved process may not be itself stochastic. The forecast skill of ensemble forecast systems has been shown to improve with these stochastic parameterizations (Ibid.). Deterministic integrations with models that include these parameterizations have  also been shown to improve climate features (see e.g. \citealt{lott12}). In general, stochastic parameterizations are expected to improve coarse-grained models of multi-scale physical systems \citep{katsoulakis03,majda11}. However, the functional form of the schemes and their parameters, which represents small-scale effects, are unknown and must be inferred from observations. The development of automatic statistical learning techniques to identify an optimal stochastic parameterization and estimate its parameters is, therefore, highly desirable.

One standard methodology to estimate physical model parameters from observations in data assimilation techniques, such as the traditional Kalman filter, is to augment the state space with the parameters \citep{jazwinski70}. This methodology has also been implemented in the ensemble-based Kalman filter (see e.g. \citealt{anderson01}). The parameters are constrained through their correlations with the observed variables.  %There are alternative approaches that do not rely on state augmentation: Du and Smith \cite{du12} used probability  forecasts to estimate mean parameters, minimizing a local skill score. Abarbanel et al. \cite{abarbanel08} proposed to estimate parameters using a model-to-observations synchronization approach designed to keep the conditional Lyapunov exponent negative and with small values.  

The collapse of the parameter posterior distribution found in both ensemble Kalman filters \citep{delsole10,ruiz13a,ruiz13b,santiti15} and particle filters \citep{west01}  is a major contention point when one is interested in estimating stochastic parameters of nonlinear dynamical models. Hereinafter, we refer as stochastic parameters to those that define the covariance of a Gaussian stochastic process  \citep{delsole10}. In other words, the sequential filters are, in principle, able to estimate deterministic physical parameters, the mean of the parameter posterior distribution, through the augmented state-space procedure, but they are unable to estimate stochastic parameters of the model,  because of the collapse of the corresponding posterior distribution.  Using the Kalman filter with the augmentation method, \cite{delsole10} proved analytically the collapse of the parameter covariance in a first-order autoregressive model.  They proposed a generalized maximum likelihood estimation using an approximate sequential method to estimate stochastic parameters. \cite{carrassi11} derived the evolution of the augmented error covariance in the extended Kalman filter using a quadratic in time approximation that  mitigates the collapse of the parameter error covariance. \cite{santiti15} proposed a particle filter blended with an ensemble Kalman filter and use a random walk model for the parameters. This technique was able to estimate stochastic parameters in the first-order autoregressive model, but a tunable parameter in the random walk model needs to be introduced. %Heald and Stark \cite{heald00} proposed a Bayesian method, along with a Laplace approximation, to estimate  model and  measurement noise in chaotic dynamical systems. The proper filtering of electron radiation belt observations requires the identification of model and measurement noise covariance matrices, \cite{podladchikova14} proposed to use the residual statistics for this purpose. Kwasniok \cite{kwasniok12} used a Kalman filter with likelihood function evaluation to determine both observational and model noise parameters; the maximization is produced through gridded two-dimensional evaluation.

The Expectation-Maximization (EM) algorithm \citep{dempster77,bishop06} is a widely used methodology to maximize the likelihood function in a broad spectrum of applications. One of the advantages of the EM algorithm is that its implementation is rather straigthforward. Wu (1983) showed that if the likelihood is smooth and unimodal, the EM algorithm converges to the unique maximum likelihood estimate. Accelerations of the EM algorithm have been proposed for its use in machine learning \citep{neal99}. Recently, it was used in an application with a highly nonlinear observation operator \citep{tandeo15}. The EM algorithm was able to estimate subgrid-scale parameters with good accuracy while standard ensemble Kalman filter techniques failed. It has also been applied to the Lorenz-63 system to estimate model error covariance \citep{dreano17}. 

In this work, we combine for stochastic parameterization identification these two independent methodologies: the ensemble Kalman filter \citep{evensen94,evensen03} for the state-estimate with maximum likelihood estimators, the EM \citep{dempster77,bishop06} and the Newton-Raphson (NR) algorithms \citep{cappe05}. The derivation of the technique is explained in detail and simple terms so that readers that are not from those communities can understand the basis of the methodologies, how they can be combined, and hopefully foresee potential applications in other geophysical systems. The learning statistical techniques are suitable  to infer the functional form and the parameter values of stochastic parameterizations in chaotic spatio-temporal dynamical systems.  They are evaluated here on a two-scale spatially extended chaotic dynamical system \citep{lorenz96} to estimate deterministic physical parameters, together with additive and multiplicative stochastic parameters. \cite{pulido16} evaluated methods based on the EnKF alone to estimate subgrid-scale parameters in a two-scale system: they showed that an offline estimation method is able to recover the functional form of the subgrid-scale parameterization, but none of the methods was able to estimate the stochastic component of the subgrid-scale effects. In the present work, the results show that the NR and EM techniques are able to uncover the functional form of the subgrid-scale parameterization while succesfully determining the stochastic parameters of the representation of  subgrid-scale effects.

This work is organized as follows. Section~2 briefly introduces the EM algorithm and derives the marginal likelihood of the data using a Bayesian perspective. The implementation of the EM  and NR likehood maximization algorithms in the context of data assimilation using the ensemble Kalman filter is also discussed. Section~3 describes the experiments which are based on the one- and two-scale Lorenz-96 systems. The former includes simple deterministic and stochastic parameterizations to represent the effects of the smaller scale to mimic the two-scale Lorenz-96 system. Section~4 focuses on the results: Section~4.1 discusses the experiments for the estimation of model noise. Section~4.2 shows the results of the estimation of deterministic and stochastic parameters in a perfect-model scenario. Section~4.3 shows the estimation experiments for an imperfect model. The conclusions are drawn in Section~5.

\section{Methodology}

\subsection{Hidden Markov model}
A hidden Markov model is defined by a stochastic nonlinear dynamical model $\mathcal{M}$ that evolves in time the hidden variables $\v x_{k-1} \in \mathbb{R}^N$, according to
\mi
\v x_k = \mathcal{M}_{\gv \Omega}(\v x_{k-1})+\gv \eta_{k}, \label{model_addq}
\mf
where $k$ stands for the time index. The  dynamical model $\mathcal{M}$ depends on a set of deterministic and stochastic physical parameters denoted by $\gv \Omega$. We assume an additive random model error,   $\gv \eta_{k}$, with covariance matrix $\v Q_k=\expec{\gv \eta_{k} \gv \eta_{k}^\transp}$. The notation $\expec{}$ stands for the expectation operator, $\mathcal{E}\left[f (x)\right]\equiv \int f(x) p(x) \ud x$ with $p$ being the probability density function of the underlying process $X$. 

The observations at time $k$, $\v y_k\in\mathbb{R}^M$, are related to the hidden variables through the observational operator $\mathcal{H}$,
\mi
\v y_k = \mathcal{H}(\v x_{k})+\gv \epsilon_{k}, \label{obs_model}
\mf
where $\gv \epsilon_{k}$ is an additive random observation error with observation error  covariance matrix $\v R_k=\expec{\gv \epsilon_{k} \gv \epsilon_{k}^\transp}$.

\vskip 4mm
\noindent {\bf Our estimation problem:} {\em Given a set of observation vectors distributed in time, $\{\v y_k,k=1,\dots,K\}$, a nonlinear stochastic dynamical model,  $\mathcal{M}$,  and a nonlinear observation operator, $\mathcal{H}$, we want to estimate the initial prior distribution $p(\v x_0)$,  the observation error covariance $\v R_k$, the model error covariance $\v Q_k$, and deterministic and stochastic physical parameters $\gv \Omega$ of $\mathcal{M}$.}

Since the EM literature also uses the term parameter for the covariances, we need to distinguish them from deterministic and stochastic model parameters in this work. We refer to the parameters of a subgrid-scale parameterization (in the physical model) as {\em physical parameters}, including deterministic and stochastic ones. While the parameters of the likelihood function are referred to as {\em statistical parameters}. These include the deterministic and stochastic physical parameters, as well as the initial prior distribution, the observation error covariance and the model error covariance.

The estimation method we derive is based on maximum likelihood estimation. Given a set of independent and identically distributed  (iid) observations from a probability density function represented by $p(\v y_{1:K}|\gv \theta)$, we seek to maximize the likelihood function $L(\v y_{1:K}; \gv \theta)$ as a function of $\gv \theta$.  We denote $\{\v y_{1}, \cdots,\v y_{K}\}$  by $\v y_{1:K}$ and  the set of statistical parameters to be estimated  by $\gv \theta$: the deterministic and stochastic physical parameters $\gv \Omega$ of the dynamical model $\mathcal{M}$ as well as observation error covariances $\v R_k$, model error covariances $\v Q_k$ and the initial prior distribution $p(\v x_0)$.  In practical applications, the statistical moments $\v R_k$, $\v Q_k$ and $\v P_0$ are usually poorly constrained. It may thus be convenient  to estimate them jointly with the physical parameters. The dynamical model is assumed to be nonlinear and to include stochastic processes represented by some of the physical parameters in $\gv \Omega$.

The estimation technique used in this work is a batch method: a set of observations taken along a time interval is used to estimate the model state trajectory that is closest to them, considering measurement and model errors with a least-square criterion to be established below. The simultaneous use of observations distributed in time is essential to capture the interplay of the several statistical parameters and physical stochastic parameters included in the estimation problem. The required minimal length $K$ for the observation window is evaluated in the numerical experiments. The estimation technique may be applied in sucessive K-windows. For stochastic parameterizations in which the parameters are sensitive to processes of different time scales, a batch method may also be required to capture the sensitivity to slow processes.
 
\subsection{Expectation-maximization algorithm}
 
The EM algorithm  maximizes the log-likelihood of observations as a function of the statistical parameters $\gv \theta$ in the presence of a hidden state $\v x_{0:K}$ \footnote{We use the notation ``$;$'', $p(\v y_{1:K}; \gv \theta)$ instead of conditioning ``$|$'' to emphasize that $\gv \theta$ is not a random variable but a parameter. NR maximization and EM are point estimation methods so that $\gv \theta$ is indeed assumed to be a  parameter \citep{cappe05}.}, 
\mi
l(\gv \theta)=\ln L(\v y_{1:K}; \gv \theta)=\ln \int p(\v x_{0:K},\v y_{1:K};\gv \theta) \ud \v x_{0:K}.\label{observationLikelihood}
\mf%= \ln p(\v y_{1:K}; \gv \theta)

An analytic form for the log-likelihood function, \reff{observationLikelihood}, can be obtained only in a few ideal cases. Furthermore, the numerical evaluation of \reff{observationLikelihood} may involve high-dimensional integration of the complete likelihood (integrand of \reff{observationLikelihood}).  Given an initial guess of the statistical parameters $\gv \theta$, the EM algorithm  maximizes  the log-likelihood of observations as a function of the statistical parameters in successive iterations without the need to evaluate the complete likelihood.

\subsubsection{The  principles}

Let us introduce in the integral \reff{observationLikelihood} an arbitrary probability density function of the hidden state, $q(\v x_{0:K})$,
\mi
l(\gv \theta)=\ln \int q(\v x_{0:K}) \frac{p(\v x_{0:K},\v y_{1:K};\gv \theta)}{q(\v x_{0:K})}\ud \v x_{0:K}.
\mf
 We assume that the support of $q(\v x_{0:K})$ contains that of $p(\v x_{0:K},\v y_{1:K};\gv \theta)$. In particular, $q(\v x_{0:K})$ may be thought as a function of a set of fixed statistical parameters $\gv \theta'$, $q(\v x_{0:K};\gv \theta')$. Using Jensen inequality a lower bound for the log-likelihood is obtained,
\mi
l(\gv \theta)\ge\int q(\v x_{0:K}) \ln \left(\frac{p(\v x_{0:K},\v y_{1:K};\gv \theta)}{q(\v x_{0:K})}\right) \ud \v x_{0:K} \equiv \mathcal{Q}(q,\theta) \label{jensen}\\
\mf

If we choose $q(\v x_{0:K})= p(\v x_{0:K}|\v y_{1:K};\gv \theta')$, the equality is satisfied in \reff{jensen}, therefore  $p(\v x_{0:K}|\v y_{1:K};\gv \theta')$ is an upper bound to $\mathcal{Q}$ and so it is the $q$ function that maximises $\mathcal{Q}(q,\gv \theta)$.

From \reff{jensen} we see that if we maximize $\mathcal{Q}(q,\gv \theta)$ over $\gv \theta$, we find a lower bound for $l(\gv \theta)$. The idea of the EM algorithm is to first find the probability density function $q$ that maximizes $\mathcal{Q}$, the conditional probability of the hidden state given the observations, and then to determine the parameter $\gv \theta$ that maximizes $\mathcal{Q}$. Hence, the EM algorithm encompasses the following steps:

{\em Expectation:} Determine the distribution $q$ that maximizes $\mathcal{Q}$. This function is easily shown to be $q^*=p(\v x_{0:K}|\v y_{1:K};\gv \theta')$ (see \reff{jensen}; \citealt{neal99}). The function $q^*$ is the conditional probability of the hidden state given the observations. In practice, this is obtained by evaluating the conditional probability at $\gv \theta'$. 

{\em Maximization:}  Determine the statistical parameters $\gv \theta^*$ that maximize $\mathcal{Q}(q^*,\gv \theta)$ over $\gv \theta$. The new estimation of the statistical parameters is denoted by $\gv \theta^*$ while the (fixed) previous estimation  by $\gv \theta'$. The expectation step is a function of these old statistical parameters $\gv \theta'$. The part of function $\mathcal{Q}$ to maximize is given by
%\mi
\begin{align}
\int p(\v x_{0:K}|\v y_{1:K};\gv \theta') \ln\left(p(\v x_{0:K},\v y_{1:K};\gv \theta)\right) \ud \v x_{0:K}\equiv\nonumber \\  \mathcal{E}\left[\ln\left(p(\v x_{0:K},\v y_{1:K};\gv \theta)\right)|\v y_{1:K}\right].\label{intermediatefn}
\end{align}
%\mf
where we use the notation $\mathcal{E}\left(f (x)|y\right)\equiv \int f(x) p(x|y) \ud x$ \citep{jazwinski70}. While the function that we want to maximize is the log-likelihood, the intermediate function \reff{intermediatefn} of the EM algorithm to maximize is the expectation of the joint distribution {\em conditioned to the observations}. %The notation we use in  \reff{intermediatefn} is defined by $\mathcal{E}\left(f (x)|y\right)\equiv \int f(x) p(x|y) \ud x$ \citep{jazwinski70}.

\subsubsection{Expectation-maximization for a hidden Markov model}

The joint distribution of a hidden Markov model using the definition of the conditional probability distribution reads
\mi
 p(\v x_{0:K}, \v y_{1:K})=p(\v x_{0:K}) p(\v y_{1:K}|\v x_{0:K}) . \label{completeDist}
\mf
The model state probability density function can be expressed as a product of the transition density from $t_k$ to $t_{k+1}$ using the definition of the conditional probability distribution and the Markov property, 
\mi
p(\v x_{0:K})=p(\v x_0) \prod_{k=1}^{K} p(\v x_{k}|\v x_{k-1}) .  \label{stateEvolDist}
\mf
The observations are mutually independent and are conditioned on the current state (see \reff{obs_model}) so that
\mi
p(\v y_{1:K}|\v x_{0:K})= \prod^{K}_{k=1} p(\v y_k|\v x_{k}) . \label{obsDist}
\mf
Then, replacing \reff{stateEvolDist} and \reff{obsDist} in \reff{completeDist} yields 
\mi
p(\v x_{0:K}, \v y_{1:K})= p(\v x_0) \prod_{k=1}^{K} p(\v x_{k}|\v x_{k-1})  p(\v y_k|\v x_{k}). \label{completeDistHMM}
\mf

If we now assume a Gaussian hidden Markov model, and that the covariances $\v R_k$ and $\v Q_k$ are constant in time, the logarithm of the joint distribution \reff{completeDistHMM} is then given by

%\begin{align*}
%\ln(p(\v x_{0:K}, \v y_{1:K})) =-\frac{(M+N)}{2} \ln(2\pi)-\frac{1}{2} \ln|\v P_0| - \frac{1}{2} (\v x_0-\overline{\v x}_0)^\transp \v P_0^{-1} (\v x_0-\overline{\v x}_0)&\nonumber\\
%-\frac{K}{2} \ln|\v Q|-\frac{1}{2} \sum_{k=1}^K (\v x_k-\mdl{\v x_{k-1}})^\transp \v Q^{-1} (\v x_k-\mdl{\v x_{k-1}}) &\nonumber\\
% -\frac{K}{2} \ln|\v R|-\frac{1}{2} \sum_{k=1}^K (\v y_k-\H{\v x_{k}})^\transp \v R^{-1} (\v y_k-\H{\v x_{k}})&.\label{jointdistribution}
%\end{align*}

\newpage
\begin{widetext}
  %\mi
  \begin{align}
\ln(p(\v x_{0:K}, \v y_{1:K})) &=-\frac{(M+N)}{2} \ln(2\pi)-\frac{1}{2} \ln|\v P_0| - \frac{1}{2} (\v x_0-\overline{\v x}_0)^\transp \v P_0^{-1} (\v x_0-\overline{\v x}_0)
-\frac{K}{2} \ln|\v Q|\nonumber\\&-\frac{1}{2} \sum_{k=1}^K (\v x_k-\mdl{\v x_{k-1}})^\transp \v Q^{-1} (\v x_k-\mdl{\v x_{k-1}}) 
 -\frac{K}{2} \ln|\v R|-\frac{1}{2} \sum_{k=1}^K (\v y_k-\H{\v x_{k}})^\transp \v R^{-1} (\v y_k-\H{\v x_{k}}).\label{jointdistribution}
 %\mf
 \end{align}
\end{widetext}

The Markov hypothesis implies that model error is not correlated in time. Otherwise, we would have cross terms in the model error summation of \reff{jointdistribution}. The assumption of a Gaussian hidden Markov model is central to derive a closed form for the statistical parameters that maximize the intermediate function. However, the dynamical model and observation operator may have nonlinear dependencies so that the Gaussian assumption is not strictly held. We therefore consider an iterative method in which each step is an approximation. In general, the method will converge through sucessive approximations. For severe nonlinear dependencies in the dynamical model, the existence of a single maximum in the log-likelihood is not guaranteed. In that case, the EM algorithm may converge to a local maximum.

We consider \reff{jointdistribution} as a function of the statistical parameters in this  Gaussian state-space model. As mentioned, the statistical parameters, which are in general denoted by $\gv \theta$, are $\overline{\v x}_0$, $\v P_0$, $\v Q$, $\v R$, and the physical parameters from $\mathcal{M}$. In this way, the log-likelihood function is written as
\mi
\l(\gv \theta)= \ln L(\gv \theta)=\ln(p(\v x_{0:K}, \v y_{1:K};\gv \theta))
\mf
In this Gaussian state-space model,  the maximum of the intermediate function in the EM algorithm, \reff{intermediatefn}, may be determined analytically from
\mia
0 &=&\nabla_{\gv \theta}\mathcal{E}\left[\ln\left(p(\v x_{0:K},\v y_{1:K};\gv \theta)\right)|\v y_{1:K}\right]\nonumber \\
&=& \int p(\v x_{0:K}|\v y_{1:K};\gv \theta') \nabla_{\gv \theta} \ln(p(\v x_{0:K}, \v y_{1:K};\gv \theta)) \, \ud \v x_{0:K} \nonumber \\
&=&\mathcal{E}\left[\nabla_{\gv \theta} \ln\left(p(\v x_{0:K},\v y_{1:K};\gv \theta)\right)|\v y_{1:K}\right]\label{Qderivative}\\
\nonumber
\mfa
Note that $\gv \theta'$  is fixed in \reff{Qderivative}. We only need to find the critical values of the statistical parameters $\v Q$ and $\v R$. The physical parameters are appended to the state, so that their model error is included in $\v Q$. The  $\overline{\v x}_0$, $\v P_0$ are at the initial time so that they are obtained as an output of the smoother which gives a Gaussian approximation of $p(\v x_k|\v y_{1:K})$ with $k=0,\cdots,K$.  The smoother equations are shown in the Appendix.

Differentiating \reff{jointdistribution} with respect to $\v Q$ and $\v R$ and applying the expectation conditioned to the observations, we can determine the root of the condition, \reff{Qderivative}, which gives the maximum of the intermediate function. The value of the model error covariance, solution of \reff{Qderivative}, is
\mi
\v Q=\frac{1}{K} \sum^K_{k=1} \esp{\left[\v x_k-\mdl{\v x_{k-1}}\right]\left[\v x_k-\mdl{\v x_{k-1}}\right]^\transp}. \label{Qest}
\mf

In the case of the observation error covariance, the solution is 
\mi
\v R=\frac{1}{K} \sum^K_{k=1} \esp{\left[\v y_k-\H{\v x_{k}}\right]\left[\v y_k-\H{\v x_{k}}\right]^\transp}. \label{Rest}
\mf

Therefore we can summarize the EM algorithm for a hidden Markov model as:

{\em Expectation: }
The required set of expectations given the observations must be evaluated at $\gv \theta_i$, $i$ being the iteration number, specifically, $\esp{\v x_k}$, $\esp{\v x_k\v x_k^\transp}$, etc. The outputs of a classical smoother are indeed $\esp{\v x_k}$, $\esp{(\v x_k-\esp{\v x_k}) (\v x_k-\esp{\v x_k})^\transp}$ which fully characterize $p(\v x_k|\v y_{1:K})$ in the Gaussian case. Hence, this expectation step involves the application of a foward filter and a backward smoother.

{\em Maximization:}
Since we assume Gaussian distributions, the optimal value of $\gv \theta_{i+1}$ can be determined analytically, which in our model are $\v Q$ and $\v R$, as derived in \reff{Qest} and \reff{Rest}. These equations are evaluated using the expectations determined in the Expectation step. 

The basic steps of this EM algorithm are depicted in Fig. \ref{EMNRalgo}a. In this work, we use an ensemble-based  Gaussian filter, the ensemble transform Kalman filter \citep{hunt07}  and the Rauch-Tung-Striebel smoother \citep{cosme12,raanes16}\footnote{In principle what is required in \reff{intermediatefn} is $p(\v x_{0:K}|\v y_{1:K})$  so that a fixed-interval smoother needs to be applied. However, it has been shown by Raanes that the Rauch-Tung-Striebel smoother and the ensemble Kalman smoother, a fixed-interval smoother, are equivalent even in the nonlinear, non-Gaussian case. %P.N. Raanes, “On the ensemble Rauch-Tung-Striebel smoother and its equivalence to the ensemble kalman smoother,” Q. J. R. Meteorol. Soc. 142, 1259–1264 (2016).
}. A short description of these methods is given in the Appendix. The empirical expectations are determined using the smoothed ensemble member states at $t_k$, $\v x^s_m(t_k)$. For instance, 
\mi
\esp{\v x_k \v x_k^\transp}=\frac{1}{N_{e}} \sum^{N_{e}}_{m=1} \v x^s_m(t_k)\v x^s_m(t_k)^\transp,
\mf
where $N_{e}$ is the number of ensemble members. Then, using these empiral expectations $\v R$ and/or $\v Q$ are computed from \reff{Qest} and/or \reff{Rest}.

The EM algorithm applied to a linear Gaussian state space model using the Kalman filter was first proposed by \cite{shumway82}. Its approximation using an ensemble of draws (Monte Carlo EM) was proposed in \cite{wei90}. It was later generalized with the extended Kalman filter and Gaussian kernels by \cite{ghahramani99}. The use of the 
EnKF and the ensemble Kalman smoother permits the extension of the EM algorithm to nonlinear high-dimensional dynamical models and nonlinear observation operators.

\subsection{Maximum likelihood estimation via Newton-Raphson}

The EM algorithm is highly versatile and can be readily implemented. However, it requires the optimal value in the maximization step to be computed analytically which limits the range of its applications. If physical parameters of a nonlinear model need to be estimated, an analytical expression for the optimal statistical parameter values may not be available. Another approach to find an estimate of the statistical parameters consists in maximizing an approximation of the likelihood function $l(\gv \theta)$ with respect to the parameters, \reff{observationLikelihood}. This maximization may be conducted using standard optimization methods \citep{cappe05}.

Following \cite{carrassi17}, the observation probability density function can  be decomposed into the product
\mi
p(\v y_{1:K};\gv \theta)=\prod^{K}_{k=1} p(\v y_k|\v y_{1:k-1};\gv \theta) ,\label{observ}
\mf
with the convention $\v y_{1:0}=\{ \varnothing \}$. In the case of sequential application  of NR maximization in successive $K$-windows,  the a priori probability distribution $p(\v x_0)$ can be taken from the previous estimation. For such a case, we leave implicit the conditioning in \reff{observ} on all the past observations, $p(\v y_{1:K};\gv \theta)=p(\v y_{1:K}|\v y_{:0};\gv \theta)$, $\v y_{:0}=\{\v y_0, \v y_{-1}, \v y_{-2}, \cdots\}$ which is called contextual evidence in \cite{carrassi17}. The times of the evidencing window, $1:K$, required for the estimation are the only ones that are kept explicit in \reff{observ}.

Replacing \reff{observ} in \reff{observationLikelihood} yields
%\mi
%l(\gv \theta) = \sum_{k=1}^K \ln p(\v y_k|\v y_{1:k-1};\gv \theta)
%=  \sum_{k=1}^K \ln\left(\int p( \v y_k|\v x_k) p( \v x_k|\v y_{1:k-1};\gv \theta) 
%\ud \v x_k\right) . \label{filterint}
%\mf
\begin{align}
l(\gv \theta) &= \sum_{k=1}^K \ln p(\v y_k|\v y_{1:k-1};\gv \theta)\nonumber\\
&=  \sum_{k=1}^K \ln\left(\int p( \v y_k|\v x_k) p( \v x_k|\v y_{1:k-1};\gv \theta) 
\ud \v x_k\right) . \label{filterint}
\end{align}

If we assume Gaussian distributions and linear dynamical and observational models, the integrand in \reff{filterint} is exactly the analysis distribution given by a Kalman filter \citep{carrassi17}. The likelihood of the observations conditioned on the state at each time is then given by
%\mi
%p(\v y_k|\v x_{k})= [(2\pi)^{M/2}|\v R|^{1/2}]^{-1} \exp\left[-\frac{1}{2}(\v y_k-\mathcal{H}(\v x_{k}))^\transp \v R^{-1} (\v y_k-\mathcal{H}(\v x_{k}))\right], \label{likObsStt}
%\mf
\begin{align}
p(\v y_k|\v x_{k})&= [(2\pi)^{M/2}|\v R|^{1/2}]^{-1} \nonumber \\ &\exp\left[-\frac{1}{2}(\v y_k-\mathcal{H}(\v x_{k}))^\transp \v R^{-1} (\v y_k-\mathcal{H}(\v x_{k}))\right], \label{likObsStt}
\end{align}
and the prior forecast distribution,
%\mi
%p(\v x_{k}|\v y_{1:k-1};\gv \theta) = [(2\pi)^{N/2}|\v P^f_k|^{1/2}]^{-1} \exp\left[-\frac{1}{2}(\v x_k - \v x^f_k)^\transp  (\v P^f_k)^{-1}(\v x_k - \v x^f_k )\right],\label{postFor}
%\mf
\begin{align}
p(\v x_{k}|\v y_{1:k-1};\gv \theta) &= [(2\pi)^{N/2}|\v P^f_k|^{1/2}]^{-1} \nonumber \\ &\exp\left[-\frac{1}{2}(\v x_k - \v x^f_k)^\transp  (\v P^f_k)^{-1}(\v x_k - \v x^f_k )\right],\label{postFor}
\end{align}
where $\v x^f_k=\mathcal{M}(\v x^a_{k-1})+\gv \eta_{k}$ is the forecast with $\v Q_k=\expec{\gv \eta_{k} \gv \eta_{k}^\transp}$,  $\v x^a_{k-1}$ is the analysis state ---filter mean state estimate--- at time $k-1$, and $\v P^f_k$ is the forecast covariance matrix of the filter.

The resulting approximation of the observation likelihood function  which is obtained replacing \reff{likObsStt} and  \reff{postFor} in \reff{filterint},  is
%\mi
%l(\gv \theta)\approx-\frac{1}{2}\sum_{k=1}^K \left[(\v y_k-\v H \v x^f_k)^\transp (\v H \v P^f_k \v H^\transp + \v R)^{-1} (\v y_k-\v H \v x^f_k)  + \ln (|\v H \v P^f_k  \v H^\transp + \v R|)\right] +C \label{innoLogLik}
%\mf 
\begin{align}
l(\gv \theta)&\approx-\frac{1}{2}\sum_{k=1}^K \left[(\v y_k-\v H \v x^f_k)^\transp (\v H \v P^f_k \v H^\transp + \v R)^{-1} \right.\nonumber \\ &\left.(\v y_k-\v H \v x^f_k)  + \ln (|\v H \v P^f_k  \v H^\transp + \v R|)\right] +C \label{innoLogLik}
\end{align}
where $C$ stands for the constants independent of $\gv \theta$ and the observational operator is assumed linear, $\mathcal{H}=\v H$. Equation \reff{innoLogLik} is exact for linear models $\mathcal{M}=\v M$, but just an approximation for nonlinear ones. As in EM, the point we made is that we expect that {\em the likelihood in the iterative method can converge through sucessive approximations}.

The evaluation of the model evidence \reff{innoLogLik} does not require the smoother. The forecasts $\v x^f_k$ in \reff{innoLogLik} are started from the analysis ---filter state estimates. In this case, the initial statistical parameters $\v x_0$ and $\v P_0$ need to be good approximations (e.g. an estimation from the previous evidencing window) or they need to be  estimated jointly to the other potentially unknown parameters $\gv \Omega$, $\v R$, and $\v Q$. Note that \reff{innoLogLik} does not depend explicitly on $\v Q$ because the forecasts $\v x^f_k$ already include the model error. The steps of the NR method are sketched in Fig. \ref{EMNRalgo}b.

For all the cases in which we can find an analytical expression for the maximization step of the EM algorithm, we can also derive a gradient of the likelihood function \citep{cappe05}. However, we  apply the NR maximization in both cases; when the EM maximization step can be derived analytically but also when it cannot. Thus, we implement a  NR maximization based on a so-called derivative-free optimization method, i.e. a method that does not require the likelihood gradient, to be described in the next section.

\renewcommand{\baselinestretch}{1.1}
% Define block styles
\tikzstyle{decision} = [diamond, shape aspect=2, draw,  
    text width=7em, text badly centered, node distance=1cm, inner sep=0pt]
\tikzstyle{title} = [rectangle, draw, fill=gray!20, 
    text width=9em, text centered, rounded corners, minimum height=3em]
\tikzstyle{line} = [draw, -latex']
\tikzstyle{cloud} = [draw, ellipse,fill=gray!20, node distance=1cm,
    minimum height=2em]
\tikzstyle{instruc} = [rectangle, draw, 
    text width=9em, text centered, rounded corners, minimum height=4em]
\tikzstyle{assign} = [rectangle, draw, node distance=1.cm, 
    text width=9em, text centered]%, minimum height=2em]
\tikzstyle{bigassign} = [rectangle, draw, 
    text width=14em, text centered]%, minimum height=4em]
\tikzstyle{io} = [trapezium, trapezium left angle=70, trapezium right angle=110, minimum width=3cm,  node distance=2cm, minimum height=1cm,text width=15em, text centered, draw=black]
\tikzstyle{module} = [rectangle, draw, 
    text width=20em, text centered, rounded corners, minimum height=3.5em]
\tikzstyle{cuadrado} = [rectangle, draw,  text centered]
\tikzstyle{bigcuadrado} = [rectangle, text width=20em, draw,  text centered]
\tikzstyle{smodule} = [rectangle, draw, 
    text width=20em, text centered, rounded corners]

\begin{figure*}
 \begin{minipage}[t]{.49\linewidth}
\resizebox {2.8in} {!} {
\begin{tikzpicture}[node distance = 1.2cm, auto]
% Nodes
%    \node [title] (title) {EM algorithm};
   \node (a1) at (3.8,1) {(a)};
   %\node[anchor=north east] {(a)};
    \node [io] (input) {Input\\[0.3em] $\v X_0^{(0)}$, $\v y_{1:K}$, $\v R^{(0)}$, $\v Q^{(0)}$};
    \node [bigassign, node distance=1.5cm, below of=input] (initk) {Iteration index: $\quad$ i=-1\\ $l(-1)=$NaN};
    \node [assign, below of=initk] (updatek) {i=i+1};
    \node [fill=gray!20, align=left, left=2.25cm,node distance=1.cm, below of=updatek] (expecTitle) {Expectation Step};
%    \node [module, below of=initk,node distance=4cm] (enkf)  {EnKF\\ Input: \, $\v X_0$, $\v y_{1:K}$, $\v Q$, $\v R$\\ Output: $\v X_{1:K}^{a}$ and  $\v X_{1:K}^{f}$};
    \node [module, below of=initk,node distance=3.05cm] (enkf)  {Filter\\[0.2em]
$\v X_{1:K}^{a}$, $\v X_{1:K}^{f}$ = EnKF($\v X_0^{(i)}$, $\v y_{1:K}$, $\v Q^{(i)}$, $\v R^{(i)}$)};
    \node [module, below of=enkf,node distance=1.8cm] (enks)  {Smoother \\[0.2em] $\v X_{0:K}^s$=RTS( $\v X_{1:K}^{a}$,  $\v X_{1:K}^{f}$) };
%    \node [module, below of=enkf] (enks)  {Ensemble RTS smoother\\ Input: \, $\v X_{1:K}^{a}$ and  $\v X_{1:K}^{f}$\\ Ouput: \, $\v X_{0:K}^s$};
    \node [bigcuadrado, below of=enks,node distance=2.2cm] (expec)  {Evaluation of expectations\\[0.3em]
$\mathcal{E}\left(\v x_k \v x_k^\transp | \v y_{1:K}\right) = \frac{1}{N_e} \sum^{N_e}_{m=1} \v x^s_m(t_k) \v x^s_m(t_k)^\transp$\\
$\mathcal{E}\left(\mdl{\v x_{k-1}} \mdl{\v x_{k-1}}^\transp | \v y_{1:K}\right) = \frac{1}{N_e} \sum^{N_e}_{m=1} \mdl{\v x^s_m(t_{k-1})} \mdl{\v x^s_m(t_{k-1})}^\transp$\\
%$\esp{\left(\v x_k-\mdl{\v x_{k-1}}\right)\left(\v x_k-\mdl{\v x_{k-1}}\right)^\transp}=\frac{1}{N_e} \sum^{N_e}_{m=1}(\v x^s_m(t_k)-\mathcal{M}(\v x^s_m(t_{k-1})) (\v (x^s_m(t_k)-\mathcal{M}(\v x^s_m(t_{k-1}))^\transp$
};
   \node[smodule, below of=enks,node distance=4.cm] (lik) {$l^{(i)}=llik(\v X_{1:K}^{f},\v y_{1:K}, \v R^{(i)}$)}; 
   \node[cuadrado,  fit= (expecTitle)(enkf) (enks) (lik) (expec),inner sep=1mm] (expectStep) {};
    \node [fill=gray!20, align=left, left=2.15cm,node distance=1.2cm, below of=lik] (maxTitle) {Maximization Step};
   \node[bigcuadrado, below of=expec,node distance=4.45cm] (QyR) {Update of $\v \theta^{(i)}$\\
$\v Q^{(i+1)}$ from Eq. (\ref{Qest})\\
%\frac{1}{K} \sum^K_{k=1} \esp{\left(\v x_k-\mdl{\v x_{k-1}}\right)\left(\v x_k-\mdl{\v x_{k-1}}\right)^\transp}$\\
$\v R^{(i+1)}$ from Eq. (\ref{Rest})\\
%=\frac{1}{K} \sum^K_{k=1} \esp{\left(\v y_k-\H{\v x_{k}}\right)\left(\v y_k-\H{\v x_{k}}\right)^\transp}$\\
$\v X_0^{(i+1)}=\v X^s(t_0)$
};
   \node[cuadrado,  fit= (maxTitle)(QyR),inner sep=1mm] (maxStep) {};
  \node[decision,node distance=2.7cm, below of=maxStep] (ifblo)  {$i \le i_{max}$ and\\ $l(i)-l(i-1) < \epsilon$ };
    \node [io, below of=ifblo] (output) {Output\\[0.3em]  $\v X_0^{(i+1)}$,  $\v R^{(i+1)}$, $\v Q^{(i+1)}$};
%   \node[bigcuadrado, below of=QyR](ifblo)  {$i>i_{max}$\\ $l(i+1)-l(i) < \epsilon$ };
    % Draw edges
%    \path [line] (title) -- (input);
    \path [line] (input) -- (initk);
    \path [line] (initk) -- (updatek);
    \path [line] (updatek) -- (enkf);
    \path [line] (enkf) -- (enks);
    \path [line] (enks) -- (expec);
    \path [line] (expec) -- (lik);
    \path [line] (lik) -- (QyR);
    \path [line] (QyR) -- (ifblo);
%    \path [line] (ifblo) -| node [near start] {yes} (updatek);
%    \path [line] (ifblo) -|  (updatek);
  \path [line] (ifblo.east) --  node{yes}++(2.3,0) |-(updatek.east);
    \path [line] (ifblo) -- node{no} (output);
%\node [title] (title) {EM algorithm};
\end{tikzpicture}
}
%\figcaption{Flowchart of the EM algorithm}\label{emAlgo}
   \end{minipage}
%   \hspace{.01\linewidth}
%
%  NR
% 
   \begin{minipage}[b]{.49\linewidth}
 \resizebox {2.8in} {!} {
\begin{tikzpicture}[node distance = 1.2cm, auto]
% Nodes
%    \node [title] (title) {EM algorithm};
   \node (a1) at (3.8,1) {(b)};
    \node [io] (input) {Input\\[0.3em] $\v X_0$, $\v y_{1:K}$, $\v R^{(0)}$, $\v Q^{(0)}$};
    \node [bigassign, node distance=1.5cm, below of=input] (initk) {Iteration index: $\quad$ i=-1\\ $l(-1)=$NaN};
    \node [assign, below of=initk] (updatek) {i=i+1};

    \node [module, below of=initk,node distance=2.05cm] (enkf)  {Filter\\[0.2em]
$\v X_{1:K}^{a}$, $\v X_{1:K}^{f}$ = EnKF($\v X_0$, $\v y_{1:K}$, $\v Q^{(i)}$, $\v R^{(i)}$)};

   \node[smodule, below of=enkf,node distance=1.5cm] (lik) {$l^{(i)}=llik(\v X_{1:K}^{f},\v y_{1:K}, \v R^{(i)}$)}; 
    \node [module, below of=lik,node distance=1.5cm] (optimization)  {Optimization\\[0.2em] $\v Q^{(i+1)}$, $\v R^{(i+1)}$ = newuoa($l^{(0:i)}, \v Q^{(0:i)},\v R^{(0:i)}$)};
  \node[decision,node distance=2.1cm, below of=optimization] (ifblo)  {$i \le i_{max}$ and\\ $l(i)-l(i-1) < \epsilon$ };
    \node [io, below of=ifblo,node distance=2.3cm] (output) {Output\\[0.3em]  $\v R^{(i+1)}$, $\v Q^{(i+1)}$};
    \path [line] (input) -- (initk);
    \path [line] (initk) -- (updatek);
    \path [line] (updatek) -- (enkf);
    \path [line] (enkf) -- (lik);
    \path [line] (optimization) -- (ifblo);
  \path [line] (ifblo.east) --  node{yes}++(2.3,0) |-(updatek.east);
    \path [line] (ifblo) -- node{no} (output);
\end{tikzpicture}
}
\caption{(a) Flowchart of the EM algorithm (left panel). (b) NR flowchart (right panel). Each column of the matrix $\v X_k$ is an ensemble member state $\v X_k\equiv\v x_{1:N_e}(t_k)$ at time $k$. Subscript $(i)$ means $i$-th iteration. A final application of the filter may be required to obtain the updated analysis state at $i+1$. The function $llik$ is the log-likelihood calculation from \reff{innoLogLik}.%For deterministic parameter estimation, the state $\v x_k$ incorporates these parameters apart from model state variables. 
}\label{EMNRalgo}
%
%%, $X^f_k=\mdl{X^a_{k-1}} + \eta_{k-1}$ with $\v Q^{(i)}\approx \eta_{k-1} \eta_{k-1}^\transp$.}\label{EMNRalgo}
\end{minipage}
\end{figure*}
\renewcommand{\baselinestretch}{1.2}

\section{Design of the numerical experiments}

A first set of numerical experiments consists of twin experiments with a perfect model in which we first generate a set of noisy observations using the model with known parameters. Then, the maximum likelihood estimators are computed using the same model with the synthetic observations. Since we know the true parameters, we can evaluate the error in the estimation and the performance of the proposed algorithms. A second set of experiments applies the method for model identification. The (imperfect) model represents the multi-scale system through a set of coarse-grained dynamical equations and an unknown stochastic physical parameterization. The model-identification experiments are imperfect model experiments in which we seek to determine the stochastic physical parameterization of the small-scale variables from observations.  In particular, the ``nature'' or true model is the two-scale Lorenz-96 model and it is used to generate the synthetic observations, while the imperfect model is the one-scale Lorenz-96 model forced by a physical parameterization which has to be indentified. This parameterization should represent the effects of small-scale variables on the large-scale variables. In this way, the coarse-grained one-scale model with a physical parameterization with tunable deterministic and stochastic parameters is adjusted to account for the (noisy) observed data. We evaluate whether the EM algorithm and the NR method are able to determine the set of optimal parameters, assuming they exist.
 
The synthetic observations are taken from the known nature integration by, see \reff{obs_model}, 
\mi \v y_k ~=~ \v H \v x_k + \gv \epsilon_k \label{linobs_ope}\mf 
with $\v H= \v I$, i.e. all the state is observed. Futhermore, we assume non-correlated observations $\v R_k=\expec{\gv \epsilon_{k} \gv \epsilon_{k}^\transp}=  \alpha_R \v I $.

\subsection{Perfect-model experiments}

In the perfect-model experiments, we use the one-scale Lorenz-96 system and a physical parameterization that represents subgrid-scale effects. The nature integration is conducted with this model and a set of ``true'' physical parameter values. These parameters characterize both deterministic and stochastic processes. By virtue of the perfect model assumption, the model used in the estimation experiments is exactly the same as the one used in the nature integration except that the physical parameter values are assumed to be unknown. Although for simplicity  we call this  ``perfect model experiment'', this experiment could be thought as a model selection experiment with parametric model error in which we know the ``perfect functional form of the dynamical equations'' but the model parameters are completely unknown and they need to be selected from noisy observations.

The equations of the one-scale Lorenz-96 model are   
\mi 
\d{X_n}{t} + X_{n-1} (X_{n-2} - X_{n+1}) + X_n ~=~   
G_n(X_n, a_0,\cdots, a_J) \ , 
\label{dpar_eq} 
\mf 
where $n= 1, \dots, N$. The domain is assumed periodic, $X_{-1} \equiv X_{N-1}$, $X_0 \equiv X_{N}$, and  $X_{N+1} \equiv X_1 $. 

We have included in the one-scale Lorenz-96 model a {\em physical parameterization} which is taken to be, 
 \mi 
G_n(X_n,~a_0,\cdots, a_2)~=~ \sum_{j=0}^2( a_j~+~\eta_j(t) ) \cdot (X_n)^j \ , 
\label{parameterization} 
\mf 
where a noise term, $\eta_j(t)$, of the form,
\mi 
\eta_j(t)~=\eta_j ~(t-\Delta t)~+~ \sigma_j~ \nu_j(t), %~ \phi~ \eta_j ~(t-\Delta t)~+~ \sigma_j~ (1-\phi^2)^{1/2} ~\nu_j(t) \ ,  
\label{AR1} 
\mf 
has been added to {\em each} deterministic parameter. Equation \reff{AR1} represents a random walk with standard deviation of the process $\sigma_j$, the stochastic parameters, and $\nu_j(t)$ is a realization of a Gaussian distribution with zero mean and unit variance. The parameterization \reff{parameterization} is assumed to represent subgrid-scale effects, i.e. effects produced by the small-scale variables to the large-scale variables \citep{wilks05}.

%\subsection{Imperfect-model experiments}
\subsection{Model-identification experiments}

In the model-identification experiments, the nature integration is conducted with the two-scale Lorenz-96  model \citep{lorenz96}. The state of this integration is taken as the true state evolution. The equations of the two-scale Lorenz-96  model,  ``true'' model, are given by $N$ equations of large-scale variables $X_n$,  
%\mi 
%\d{X_n}{t} + X_{n-1} (X_{n-2} - X_{n+1}) + X_n ~=~ F-\frac{h~c}{b} \sum_{j=N_S/N(n-1)+1}^{n N_S/N} Y_j \ ;  
%\label{ls_eq} 
%\mf 
\begin{align}
\d{X_n}{t} + X_{n-1} &(X_{n-2} - X_{n+1}) + X_n = \nonumber \\&F-\frac{h~c}{b} \sum_{j=N_S/N(n-1)+1}^{n N_S/N} Y_j \ ;  
\label{ls_eq}
\end{align}
with $n= 1, \dots, N$; and  $N_S$ equations of small-scale variables $Y_m$, given by 
%\mi 
%\d{Y_m}{t} + c~b~ Y_{m+1} (Y_{m+2} - Y_{m-1}) + c~Y_m ~=~ \frac{h~c}{b}~ X_{\mathrm{int}[(m-1)/N_S/N]+1} \ ,  
%\label{ss_eq} 
%\mf 
\begin{align}
\d{Y_m}{t} + c~b~ Y_{m+1} &(Y_{m+2} - Y_{m-1}) + c~Y_m ~=~\nonumber \\ & \frac{h~c}{b}~ X_{\mathrm{int}[(m-1)/N_S/N]+1} \ ,  
\label{ss_eq} 
\end{align}
where $m= 1, \dots, N_S$.  The two set of equations, \reff{ls_eq}  and \reff{ss_eq}, are assumed to be defined on  a periodic domain,  $X_{-1} \equiv X_{N-1}$, $X_0 \equiv X_{N}$, $X_{N+1} \equiv X_1 $, and  $Y_0 \equiv Y_{N_S}$, $ Y_{N_S+1} \equiv Y_{1}$,  $ Y_{N_S+2} \equiv Y_{2}$. 

The imperfect model used in the model-identification experiments is the one-scale Lorenz-96 model \reff{dpar_eq} with a parameterization \reff{parameterization} meant to represent small-scale effects (right-hand side of \reff{ls_eq}).

\subsection{Numerical experiment details}

As used in previous works (see e.g., \citealt{wilks05,pulido16}), we set $N=8$ and $M=256$ for the large- and small-scale variables respectively. The constants are set to the standard values $b=10$, $c=10$ and $h=1$. The ordinary differential equations \reff{ls_eq}-\reff{ss_eq} are solved by a fourth-order Runge-Kutta algorithm. The time step is set to $dt=0.001$ for integrating \reff{ls_eq} and  \reff{ss_eq}.

For the model-identification experiments, we aim to mimic the dynamics of the large-scale equations of the two-scale Lorenz-96 system with the one-scale Lorenz-96 system \reff{dpar_eq} forced by a physical parameterization \reff{parameterization}. In other words, our nature is the two-scale model, while our imperfect coarse-grained model is the forced one-scale model. For this reason, we take  8 variables for the one-scale Lorenz-96 model for the perfect-model experiments (as the number of large-scale variables in the model-identification experiments). Equations \reff{dpar_eq} are also solved by a fourth-order Runge-Kutta algorithm. The time step is also set to $dt=0.001$.

The EnKF implementation we use is the ensemble transform Kalman filter \citep{hunt07} without localization. A short description of the ensemble transform Kalman filter is given in the Appendix. The time interval between observations (cycle) is 0.05 (an elapsed time of 0.2 represents about 1 day in the real atmosphere considering the error growth rates; Lorenz, 1996). The number of ensemble members is set to $N_{e}=50$. The number of assimilation cycles (observation times) is $K=500$. This is the ``evidencing window'' \citep{carrassi17} in which we seek for the optimal statistical parameters.  The measurement variance error is set to $\alpha_R=0.5$ except otherwise stated. We do not use any inflation factor, since  the model error covariance matrix is estimated.

The optimization method used in the  NR maximization is  ``newuoa'' \citep{powell06}. This is an unconstrained minimization algorithm which does not require derivatives. It is suitable for control spaces of about a few hundred dimensions. This derivative-free method could eventually permit to extend the  NR maximization method to cases in which the state evolution \reff{model_addq} incorporates a non-additive model error.

\begin{figure*}
\includegraphics[width=5.5in]{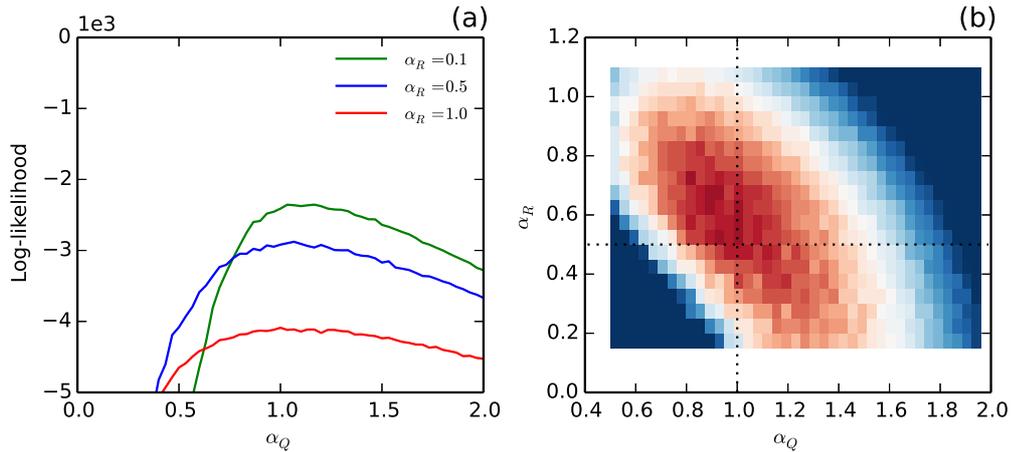} 
\vspace{-.5cm}
\caption{Log-likelihood function as a function of  (a) model noise for three true observational noise values, $\alpha_R^{t}=0.1, \, 0.5, \, 1.0$;  and as a function of (b) model noise ($\alpha_Q$) and observational noise ($\alpha_R$) for a case with $\alpha_Q^{t}=1.0$ and $\alpha_R^{t}=0.5$. Darker red shading represents larger log-likelihood.} 
\label{geom_01} 
\end{figure*}

\section{Results}

\subsection{Perfect-model experiment: Estimation of model noise parameters}

The nature integration is obtained from the one-scale Lorenz-96 model \reff{dpar_eq} with a constant forcing of $a_0=17$ without higher orders in the parameterization; in other words a one-scale Lorenz-96 model with an external forcing of $F=17$. Information quantifiers 
 show that for an external forcing of $F=17$, the Lorenz-96 model is in a chaotic regime with maximal statistical complexity  \citep{pulido17}. The true model noise covariance is defined by $ \v Q^t= \alpha_Q^{t} \v I $ with $\alpha_Q^{t}=1.0$ (true parameter values are denoted by a $t$ superscript). The observations are taken from the nature integration and perturbed using \reff{linobs_ope}.

A first experiment examines the log-likelihood  \reff{innoLogLik} as a function of $\alpha_Q$ for different true measurement errors, $\alpha_R^{t}=0.1, 0.5, 1.0$ (Fig.~\ref{geom_01}a). A relatively smooth function is found with a well-defined maximum.  The function is better conditioned for the experiments with smaller observational noise, $\alpha_R$. Figure~\ref{geom_01}b shows the log-likelihood as a function of $\alpha_Q$ and $\alpha_R$. The darkest shading is  around $(\alpha_Q, \alpha_R)\approx (1.0, 0.5)$. However, note that because of the asymmetric shape of the log-likelihood function (Fig.~\ref{geom_01}a), the darker red region is shifted toward higher  $\alpha_Q$ and $\alpha_R$ values. The up-left bottom-right orientation of the likelihood pattern in the plane  $\alpha_Q$ and $\alpha_R$ reveals a correlation between them: the larger  $\alpha_Q$, the smaller $\alpha_R$ for the local maximal likelihood. 

\begin{figure*}
\begin{center}
\includegraphics[width=4.7in]{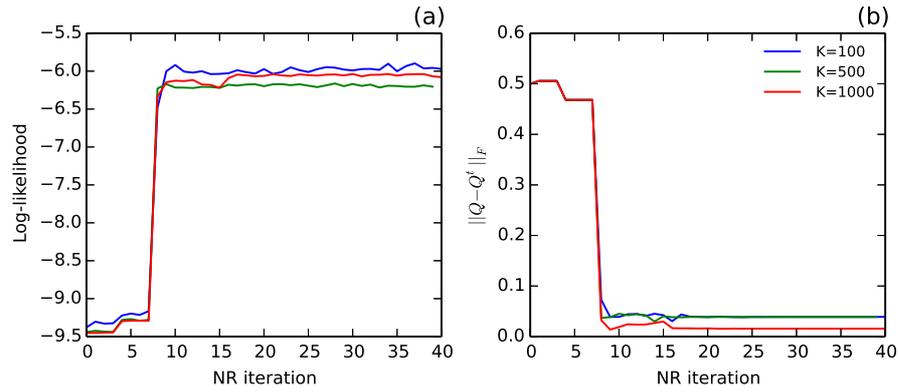} 
\end{center}
%\vskip -0.4cm 
\vspace{-.5cm}
\caption{Convergence of the NR maximization as a function of the iteration of the outer loop for different evidencing window lengths. (a) Log-likelihood function. (b) Frobenius norm of the model noise estimation error.} %(c) RMSE of the analysis state (per cycle at the corresponding iteration).} 
\label{convNR_01} 
\end{figure*}

We conducted a second experiment using the same observations but the estimation of model noise covariance matrix is performed through the NR method. The control space is of 8x8=64 dimensions, i.e. the full $\v Q$ model error covariance matrix is estimated (note that $N=8$ is the model state dimension). Figure~\ref{convNR_01}a depicts the  convergence of the log-likelihood function in three experiments with evidencing window $K=100$, $500$ and $1000$. The Frobenius norm of the error in the estimated  model noise covariance matrix, i.e. $\|\v Q-\v Q^t\|_F=\sqrt{\sum_{ij}\left(Q_{ij}-Q_{ij}^t\right)^2}$, is shown in Fig.~\ref{convNR_01}b. As the number of cycles used in a single batch process increases, the estimation error diminishes. 

\begin{figure*}
\includegraphics[width=4.7in]{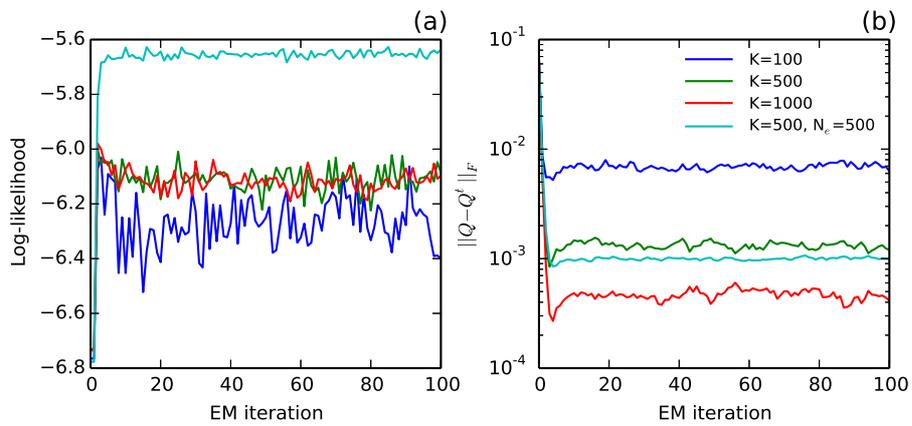} 
%\vskip -0.4cm 
\vspace{-.5cm}
\caption{Convergence of the EM algorithm as a function of the iteration  for different observation time lengths (evidencing window). An experiment with $N_e=500$ ensemble members and $K=500$ is also shown. (a) Log-likelihood function. (b) The Frobenius norm of the model noise estimation error.}% (c) RMSE of the analisis state (per cycle).} 
\label{convEM_01} 
\end{figure*}

\begin{figure*}
%\begin{center}
\includegraphics[width=4.7in]{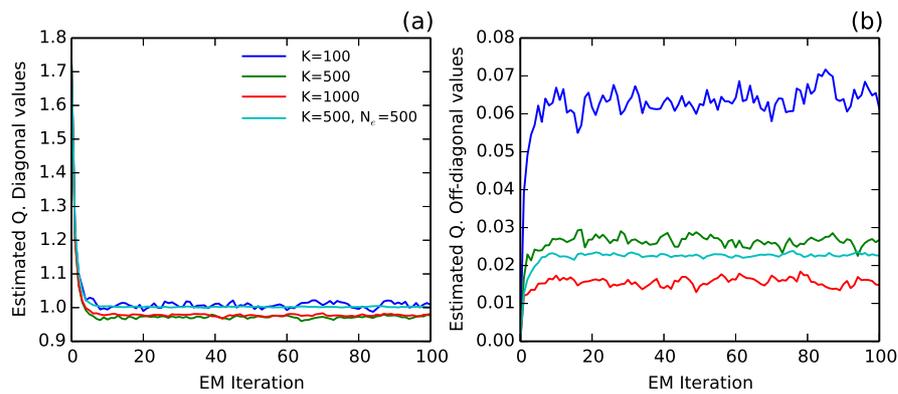} 
%\end{center}
%\vskip -0.4cm 
\vspace{-.5cm}
\caption{Estimated model noise  as a function of the iteration in the EM algorithm. (a) Mean diagonal model noise (true value is 1.0). (b) Mean off-diagonal absolute model noise value (true value is 0.0). } 
\label{convEM_02} 
\end{figure*}

The convergence of the EM algorithm applied for the estimation of model noise covariance matrix only (8x8=64 dimensions) is shown in Fig.~\ref{convEM_01}. This work is focused on the estimation of physical parameters so that the observation error covariance matrix is assumed to be known. The method would allow to estimate it jointly through \reff{Rest}, however this is beyond the main aim of this work. This is similar to the previous experiment, using the EM instead of  the NR method. In 10 iterations, the EM algorithm achieves a reasonable estimation, which is not further improved for larger number of iterations. The obtained log-likelihood value  is rather similar to the NR method. The noise in the log-likelihood function diminishes with longer evidencing windows.  Comparing the standard $N_e=50$ experiments with $N_e=500$ in Fig.~\ref{convEM_01}a, the noise also diminishes by increasing the number of ensemble members. Increasing the number of members does not appear to impact on the estimation of off-diagonal values, but it does so on the diagonal stochastic parameter values (Fig.~\ref{convEM_02}a and b). The error in the estimates is about 7\% in both diagonal and off-diagonal terms of the model noise covariance matrix for $K=100$, and lower than $2\%$ for the $K=1000$ cycles case (Fig.~\ref{convEM_02}).

%\hskip -0.7cm

%\hskip -0.7cm

%\hskip -0.7cm

\subsection{Perfect-model experiment: Estimation of deterministic and stochastic parameters}

A second set of perfect-model experiments evaluates the estimation of deterministic and stochastic parameters from a physical parameterization. The model used to generate the synthetic observations is \reff{dpar_eq} with the physical parameterization \reff{parameterization}. The deterministic parameters to conduct the nature integration are fixed to $a_0^t=17.0$, $a_1^t=-1.15$, and $a_2^t=0.04$ and the model error variance in each parameter is set to $\sigma_{0}^t=0.5,$ $\sigma_1^t=0.05,$ and $\sigma_2^t=0.002$ respectively.  The true parameters are governed by a stochastic process \reff{AR1}. This set of deterministic parameters is a representative physical quadratic polynomial parameterization, which closely resembles the dynamical regime  of a two-scale Lorenz-96 model with $F=18$ \citep{pulido17}. The observational noise is set to $\alpha_R=0.5$. An augmented state space of 11 dimensions is used, which is composed by appending to the 8 model variables the 3 physical parameters. The evolution of the augmented state is represented by \reff{model_addq} for the state vector component and a random walk for the parameters. The EM algorithm is then used to estimate the additive augmented state model error $\v Q$ which is an 11x11 covariance matrix. Therefore, the smoother recursion gives an estimate of both the state variables and deterministic parameters. The recursion formula for  the model error covariance matrix (and the parameter covariance submatrix) is given by \reff{Qest}.

\begin{figure*}
\includegraphics[width=4.7in]{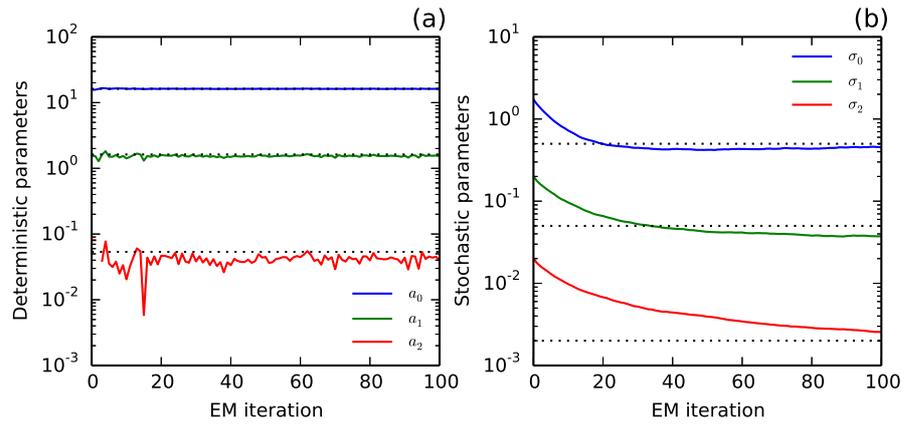} 
\vspace{-.5cm}
\caption{(a) Estimated mean deterministic parameters, $a_i$, as a function of the EM iterations for the perfect-model parameter experiment. (b)  Estimated stochastic parameters, $\sigma_i$.}\label{sparEM}
\end{figure*}

Figure~\ref{sparEM}a shows the estimation of the mean deterministic parameters as a function of the EM iterations.  The estimation of the deterministic parameters is rather accurate; $a_2$ has a small true value and it presents the lowest sensitivity. The estimation of the stochastic parameters by the EM algorithm converges rather precisely to the true stochastic parameters (Fig.~\ref{sparEM}b). The convergence requires of about 80 iterations. The estimated model error for the state variables is in the order of $5 \times 10^{-2}$. This represents the additive inflation needed by the filter for an optimal convergence. It establishes a lower threshold for the estimation of additive stochastic parameters.

A similar experiment was conducted with NR maximization for the same synthetic observations. A scaling of $S_\sigma=(1,10,100)$ was included in the optimization to increase the condition number. A good convergence was obtained with the optimization algorithm. The estimated optimal parameter values are $\sigma_0=0.38$  $\sigma_1=0.060$   $\sigma_2=0.0025$ for which the log-likelihood is $l=-491$. The estimation is reasonable with a relative error of about 25\%. 

%\subsection{Imperfect-model experiment: Estimation of the deterministic and stochastic parameters}
\subsection{Model-identification experiment: Estimation of the deterministic and stochastic parameters}

\begin{figure*}
\includegraphics[width=6.5in]{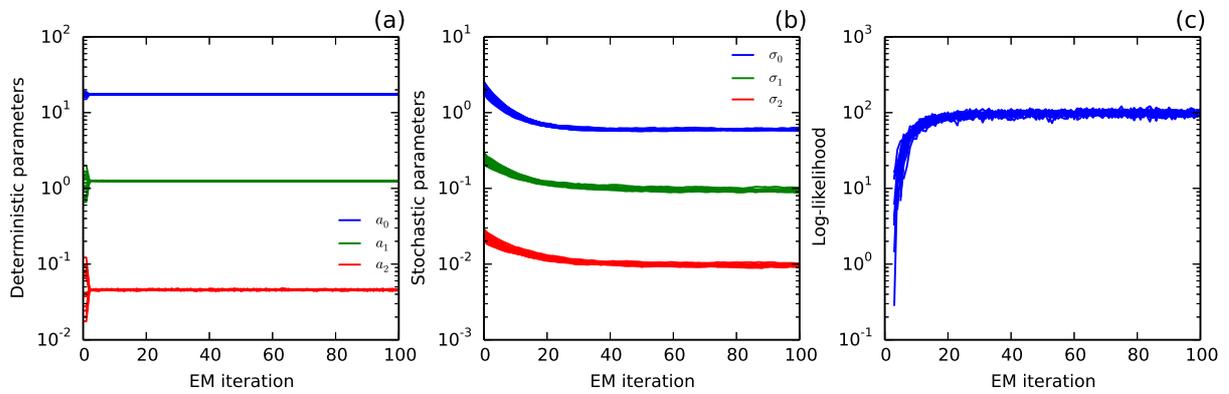} 
\vspace{-.5cm}
\caption{(a) Estimated deterministic parameters as a function of the EM iterations for the model-identification experiment. Twenty experiments with random initial deterministic and stochastic parameters are shown. (b) Estimated stochastic parameters. (c) Log-likelihood function.}\label{impEM}
\end{figure*}

As a proof-of-concept model-identification experiment, we now use synthetic observations with an additive observational noise of $\alpha_R=0.5$ taken from the nature integration of the two-scale Lorenz-96 model with $F=18$. On the other hand,  the one-scale  Lorenz-96 model is used in the ensemble Kalman filter with a physical parameterization that includes the quadratic polynomial function, \reff{parameterization}, and the stochastic process \reff{AR1}. The deterministic parameters are estimated through an augmented state space while the stochastic parameters are optimized via the algorithm for the maximization of the log-likelihood function. The model error covariance estimation is constrained for these experiments to the three stochastic parameters alone. Figure~\ref{impEM}a shows the estimated deterministic parameters as a function of the EM iterations. Twenty experiments with different initial deterministic parameters and initial stochastic parameter values were conducted. The deterministic parameter estimation does not manifest a significant sensitivity to the stochastic parameter values. The mean estimated values are $a_0=17.3$,~$a_1=-1.25$ and $a_3=0.0046$. Note that the deterministic parameter values estimated with information quantifiers in \cite{pulido17} for the two-scale Lorenz-96 with $F=18$ are $(a_0,a_1,a_2)=(17.27,-1.15,0.037)$. Figure~\ref{impEM}b depicts the convergence of the stochastic parameters. The mean of the optimal stochastic parameter values are $\sigma_0=0.60$,~$\sigma_1=0.094$ and $\sigma_2=0.0096$ with the log-likelihood value being 98.8 (single realization). The convergence of the log-likelihood is shown in Fig.~\ref{impEM}c. 

\begin{figure*}
\includegraphics[width=4.7in]{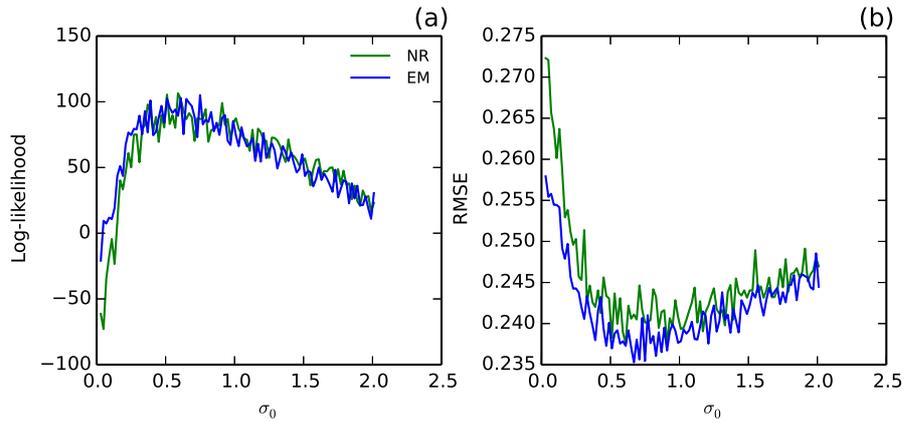} 
\vspace{-.5cm}
\figcaption{(a) Log-likelihood as a function of the $\sigma_0$ parameter at the $\sigma_1$ and $\sigma_2$ optimal values for the NR estimation (green curve) and with the optimal values for the EM estimation (blue curve) for the imperfect-model experiment. (b) Analysis RMSE as a function of the $\sigma_0$ parameter.}\label{impNR}
\end{figure*}

NR maximization is applied to the same set of synthetic observations. The mean estimated deterministic and stochastic parameters are $(a_0,a_1,a_2)=(17.2,-1.24,0.0047)$ and $(\sigma_0,\,\sigma_1,\,\sigma_2)=(0.59, 0.053, 0.0064)$ from 20 optimizations. As in the EM experiment, only the three  stochastic parameters were estimated as statistical parameters. Preliminary experiments with the full augmented  model error covariance gave smaller estimated $\sigma_0$ values and nonnegligible model error variance (not shown).
 The log-likelihood function (Fig.~\ref{impNR}a) and the  analysis root-mean-square error (RMSE, Fig.~\ref{impNR}b)  are shown as a function of $\sigma_0$ at the  $\sigma_1$ and $\sigma_2$ optimal values given by the Newton-Rapshon method (green curve) and at the  $\sigma_1$ and $\sigma_2$ optimal values given by the EM algorithm (blue curve). The log-likelihood values are indistinguishable. A slightly smaller  analysis RMSE  is obtained for the EM algorithm (Fig.~\ref{impNR}b), which is likely related to the improvement with the iterations of the initial prior distribution in the EM algorithm, while this distribution is fixed in the NR method.

\begin{figure*}
\includegraphics[width=6.5in]{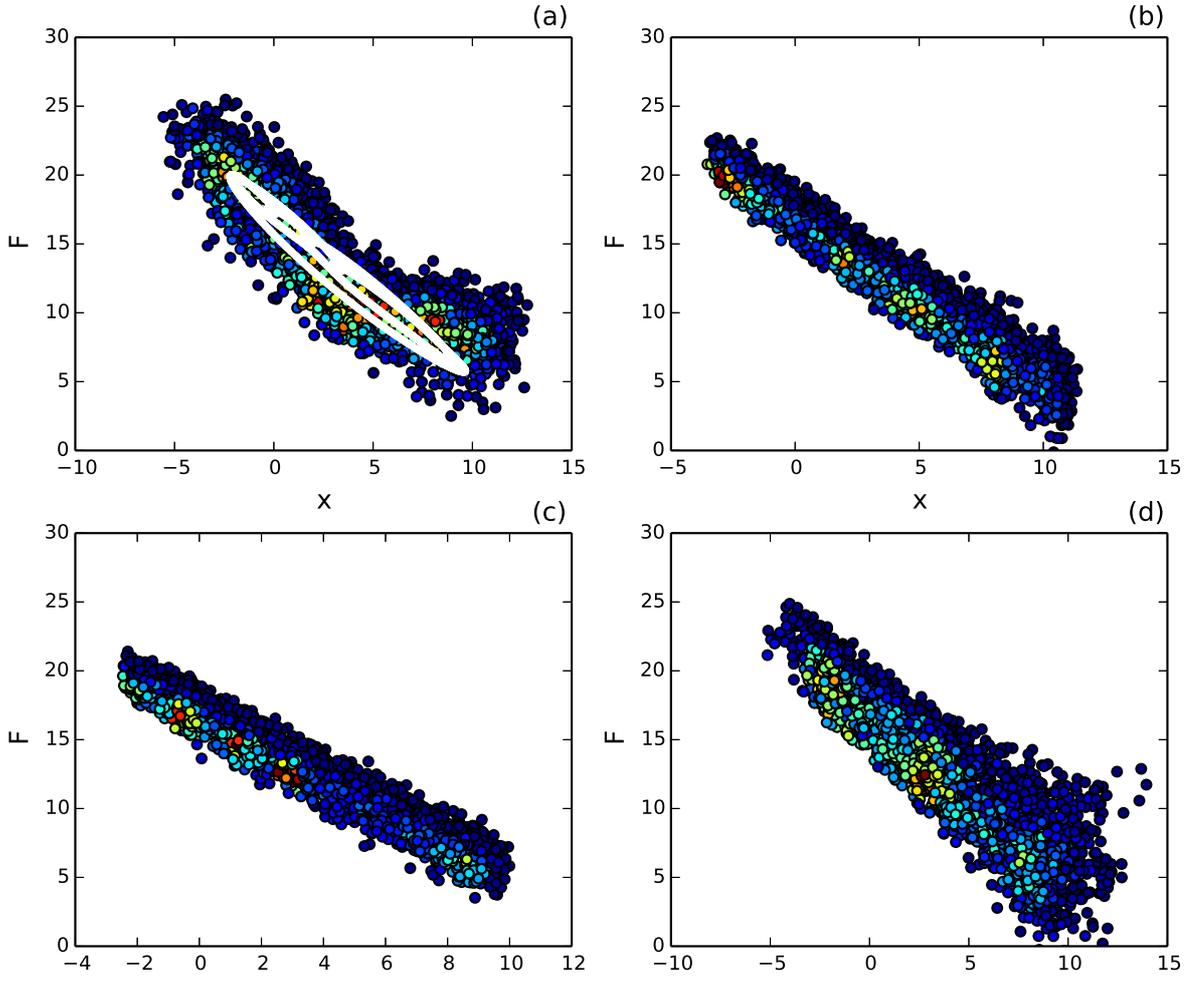} 
\vspace{-2.5cm}
\caption{(a) Scatterplot of the true small-scale effects in the two-scale Lorenz-96 model as a function of a large-scale variable (coloured dots) and scatterplot of the deterministic parameterization with optimal parameters (white dots). (b) Scatterplot from the stochastic paramerization with optimal parameters obtained with the EM algorithm and (c) with the NR method. (d) Scatterplot given by a constrained random walk with optimal EM parameters.}\label{scatterPlot1}
\end{figure*}

Long integrations ($10^6$ time cycles) of the nature model and the identified coarse-grained models were conducted to evaluate the parameterizations. The true effects of the small-scale variables on a large-scale variable from the two-scale Lorenz-96 model are shown in Fig. \ref{scatterPlot1}  as a function of the large-scale variable. This true scatterplot is obtained by evaluating the right-hand side of \reff{ls_eq}. The deterministic quadratic parameterization with the optimal parameters from the EnKF  is also represented in  Fig. \ref{scatterPlot1}(a). A poor representation of the functional form and variability is obtained. Figure \ref{scatterPlot1}(b) shows the scatterplot with a stochastic parameterization which stochastic parameters are the ones estimated with EM algorithm, while Fig. \ref{scatterPlot1}(c) shows it for the stochastic parameters estimated with the NR method. The two methods, NR and EM, give scatterplots of the parameterization which are almost indistinguishable and improve the small-scale representation with respect to the deterministic parameterization. Figure \ref{scatterPlot1}(d) shows the scatterplot resulting from the quadratic parameterization using a random walk for the parameters set to the estimated values with the EM algorithm. The values of the parameters are limited to the $a_i \pm 4 \sigma_i$ range. The parameter values need to be constrained, because for these long free simulations, some parameter values given by the random walk produce numerical instabilities in the Lorenz-96 model \citep{pulido16}. The stochastic parameterization which was identified by the statistical learning technique improves substantially the functional form of the effects  of the small-scale variables. Using a constrained random walk appears to give the best simulation.

%\begin{figure}
%\includegraphics[width=6.5in]{scatterF18color.eps} 
%\caption{Alternative Figure with colors to indicate frequency (from histogram).}\label{scatterPlot2}
%\end{figure}

\section{Conclusions}
Two methods, the EnKF-EM and EnKF-NR, have been introduced to characterize physical parameterizations in stochastic nonlinear multi-scale dynamical systems from noisy observations, which include the estimation of deterministic and stochastic parameters. Both methods determine the maximum of the observation likelihood --maximum of the model evidence-- in a time interval in which a set of spatio-temporally distributed observations are available. They use the ensemble Kalman filter to combine observations with model predictions. The methods are first evaluated in a controlled model experiment in which the true parameters are known and then, in the two-scale Lorenz-96 dynamics which is represented with a stochastic coarse-grained model. The methods do not require neither inflation factors nor any other tunable parameters. The performance of the methods is excellent, even in the presence of moderate observational noise.

The estimation based on the expectation-maximization algorithm gives very promising results in these medium-sized experiments ($\approx$100 parameters). About 50 iterations are needed to achieve an estimation error lower than 10\%  using 100 observation times. Using a longer observation time inverval, the accuracy is improved.
The estimation of stochastic parameters included the case of additive, i.e. $a_0$, and multiplicative parameters, i.e. $a_1 X_n$ and $a_2 X^2_n$. 
The number of ensemble members has a strong impact  on the stochastic parameter variance, while the length of the observation time interval appears to have a stronger impact on the stochastic parameter correlations.

The estimation based on the NR method also presents good convergence for the perfect-model experiment with an additive stochastic parameter. For the more realistic model-identification experiments, the model evidence presents some noise which may affect the convergence. For higher dimensional problems, optimization algorithms that use the gradient of the likelihood to the statistical parameters need to be implemented. Moreover, the use of simulated annealing or other stochastic gradient optimization techniques suitable for noisy cost functions would be required. 

Both estimation methods can be applied to a set of different dynamical models to address which one is more reliable given a set of noisy observations; the so called ``model selection'' problem. A comparison of the likelihood from the different models with the optimal parameters gives a measure of the model fidelity to the observations. \cite{majda11} seeked to improve imperfect models by adding stochastic forcing and used a measure from information theory that gives the closest model distribution to the observed probability distribution. The model-identification experiments in the current work can be viewed as pursuing a similar objective, stochastic processes are added to the physical parameterization to improve the model representation of the unresolved processes. A sequential Monte Carlo filter is used between observations so that their error is accounted in the estimation. In both cases, the methodologies are based on Gaussian assumptions.

\cite{hannart16} proposed to apply the observation likelihood function, model evidence, that results from assimilating a set of observations, for the detection and attribution of climate change. They suggest to evaluate the likelihood in two possible model configurations, one with the current anthropogenic forcing scenario (factual world) and one with the preindustrial forcing scenario (contrafactual world). If the evidencing window where the observations are located includes, for instance, an extreme event then  one could determine the fraction of attributable risk as the fraction of the change in the observation likelihood of the extreme event which is attributable to the anthropogenic forcing.

The increase of data availability in many areas has fostered the number of applications of the  ensemble Kalman filter. In particular, it has been used for influenza forecasting \citep{shaman13} and for determining a neural network structure \citep{hamilton13}. The increase in spatial and temporal resolution of data offers great opportunities for understanding multi-scale strongly-coupled systems such as atmospheric and oceanic dynamics. This has lead to the proposal of purely data-driven modeling which uses past observations to reconstruct the dynamics through the ensemble Kalman filter without a dynamical model \citep{hamilton16,lguensat17}. The use of automatic statistical learning techniques that can use measurements for improvement of multi-scale models is also a promising venue. Following this recent stream of research, in this work we propose the coupling of the EM algorithm and NR method with the ensemble Kalman filter which may be applicable to a wide range of multi-scale systems to improve the representation of the complex interactions between different scales.

%Following this recent stream of research,
%The present contribution is along this recent stream of research. The proposed coupling of the EM algorithm and NR method with the ensemble Kalman filter may be applicable to a wide range of multi-scale systems to improve the representation of the complex interactions between different scales.

%\begin{acknowledgments}
\vskip 3mm
{\bf Acknowledgments}
\vskip 3mm

The authors wish to acknowledge the members of the DADA CNRS team for insightful discussions, in particular Alexis Hannart, Michael Ghil and Juan Ruiz. A. Carrassi has been funded by the Nordic Center of Excellence EmblA of the Nordic Countries Research Council, NordForsk, and by the project REDDA of the Norwegian Research Council. M. Lucini and M. Pulido have been funded by PICT2015-2368 grant. Cerea is a member of Institut Pierre-Simon Laplace (IPSL).
%\end{acknowledgments}

\appendix
\section{Ensemble Kalman filter and smoother}

The ensemble Kalman filter determines  the probability density function of a dynamical model conditioned to a set of past observations, i.e. $p(\v x_k|\v y_{1:k})$, based on the Gaussian assumption. The mean and covariances are represented by a set of possible states, called {\em ensemble members}. Let us assume that the a priori ensemble members at time $k$ are $\v x^{f}_{1:N_e}(t_k)$, so that the empirical mean and covariance of the a priori hidden state are 
%\mi
%\mean{\v x}^f(t_k)=\frac{1}{N_e} \sum_{m=1}^{N_e} \v x^{f}_m(t_k), \quad \v P^f(t_k)=\frac{1}{N_e-1} \v X^f(t_k) [\v X^f(t_k)]^\transp, 
%\mf
\begin{align}
& \mean{\v x}^f(t_k)=\frac{1}{N_e} \sum_{m=1}^{N_e} \v x^{f}_m(t_k), \nonumber\\& \v P^f(t_k)=\frac{1}{N_e-1} \v X^f(t_k) [\v X^f(t_k)]^\transp, 
\end{align}
where $\v X^f(t_k)$ is a matrix with the ensemble member perturbations, $\v x^{f}_m(t_k)-\mean{\v x}^f(t_k)$, as the $m$-ith column.

To obtain the estimated hidden state, called {\em analysis state}, the observations are combined statistically with the a priori model state using the Kalman filter equations. In the case of the ensemble transformed Kalman filter \citep{hunt07}, the analysis state is a linear combination of the $N_e$ ensemble member perturbations,
\mi
\mean{\v x}^a=\mean{\v x}^f + \v X^f \mean{\v w}^a, \quad \v P^a = \v X^f \tilde{\v P}^a (\v X^f)^\transp. \label{xaETKF}
\mf
%\begin{align}
%&\mean{\v x}^a=\mean{\v x}^f + \v X^f \mean{\v w}^a, \nonumber \\&\v P^a = \v X^f \tilde{\v P}^a (\v X^f)^\transp. \label{xaETKF}
%\end{align}
The optimal ensemble member weights $\mean{\v w}^a$ are obtained considering the distance between the projection of member states to the observational space, $\v y^f_m\equiv \mathcal{H}(\v x^f_m)$,  and observations $\v y$. These weights and the analysis covariance matrix in the perturbation space are 
%\mi 
%\mean{\v w}^a = \tilde{\v P}^a (\v Y^f)^\transp \v R^{-1} [\v y-\mean{\v y}^f],
%\quad \tilde{\v P}^a = [(N_e-1) \v I + (\v Y^f)^\transp \v R^{-1} \v Y^f ]^{-1} .\label{waETKF}
%\mf 
\begin{align}
&\mean{\v w}^a = \tilde{\v P}^a (\v Y^f)^\transp \v R^{-1} [\v y-\mean{\v y}^f],\nonumber\\ &
\tilde{\v P}^a = [(N_e-1) \v I + (\v Y^f)^\transp \v R^{-1} \v Y^f ]^{-1} .\label{waETKF}
\end{align}
All the quantities in   \reff{xaETKF} and \reff{waETKF}  are at time $t_k$ so that the time dependence is omitted for clarity. A detailed derivation of \reff{xaETKF} and \reff{waETKF} and a thorough description of the ensemble transformed Kalman filter and its numerical implementation can be found in \cite{hunt07}.

To determine each ensemble member of the analysis state, the ensemble transformed Kalman filter uses the square root of the analysis covariance matrix, thus it belongs to the so-called square-root filters,
\mi
\v x^a_m= \mean{\v x}^f + \v X^f \v w^a_m
\mf
where the perturbations of $\v w^a_m$ are the columns of $\v W^a=[(N_e-1) \tilde{\v P}^a]^{1/2}$.

The analysis state is evolved to the time of the next available observation $t_{k+1}$ through the dynamical model equations which give the a priori or {\em forecasted} state,
\mi
\v x^{f}_m(t_{k+1}) = \mathcal{M}(\v x^a_m(t_k)).
\mf
%a new analysis state is then obtained applying ...

The smoother  determines the probability density function of a dynamical model conditioned to a set of past and future observations, i.e. $p(\v x_k|\v y_{1:K})$, based on the Gaussian assumption. Applying the Rauch-Tung-Striebel smoother retrospective formula to each ensemble member \citep{cosme12}, 
\mi
\v x_m^s(t_k)=\v x_m^a(t_k)+ \v K^s(t_k) [\v x^s_m(t_{k+1})-\v x^f_m(t_{k+1})], 
\mf
where $\v K^s(t_k)= \v P^a(t_{k}) \v M^\transp_{k\rightarrow k+1} [\v P^f(t_{k+1})]^{-1}$, and $\v M_{k\rightarrow k+1}$ being the linear tangent model. For the application of the smoother in conjunction with the ensemble transformed Kalman filter, the smoother gain is reexpressed as
\mi
\v K^s(t_k)=\v X^f(t_{k}) \v W^a [\v X^f(t_{k+1})]^{\dagger}. \label{smootherGain}
\mf
In practice, the peusdo-inversion of the forecast state perturbation matrix $\v X^f$ required in  \reff{smootherGain} is conducted through singular value decomposition.
\bibliographystyle{apalike}
\bibliography{ref}

%\end{thebibliography}%

\end{document}